\documentclass[11pt]{article}

\usepackage[margin=1in]{geometry}
\usepackage{amsmath,amssymb}
\usepackage{graphicx}
\usepackage{booktabs}
\usepackage[hidelinks]{hyperref}

\graphicspath{{./}}

\newcommand{\lstar}{\ell^{*}}
\newcommand{\mueff}{\mu_{\mathrm{eff}}}
\newcommand{\aeff}{\alpha_{\mathrm{eff}}}
\newcommand{\ns}{n_{s}}

\title{First-return statistics and the point-spread function of a\\
       Henyey--Greenstein random walk: a far-field stable law}
\author{Claude Zeller \and Robert Cordery}
\date{28 August 2026}

\begin{document}
\maketitle

\begin{abstract}
A particle launched normally into a half space, scattering by the
Henyey--Greenstein (HG) phase function, exits after a first-return time whose
count of flights $n$ approaches the distribution-free Sparre~Andersen exponent,
$f(n)\propto n^{-3/2}$, with an amplitude fixed by the launch rather than by the
theorem. Conditioned on $n$, and for $n$ past a
correlation-set transient, the lateral exit position is Gaussian with variance
proportional to $n$---a scaling we test directly, finding it accurate to a few
percent for $n\gtrsim100$ and breaking in the ballistic near field. Subordinating
that Gaussian displacement to the one-sided $\tfrac12$-stable first-return time
gives, in the diffusion limit, an isotropic $1$-stable scaling limit: the far tail
of the point-spread function (PSF) for a normally incident pencil beam is the
two-dimensional Cauchy tail $p(r)\propto r^{-3}$, the large-radius limit of the
half-space Poisson kernel. The index $\alpha=1$ is the product of the central
limit theorem and the distribution-free first-return exponent, so no phase
function admitting the same diffusion limit---finite step variance, short-range
angular correlation---can move it.
Absorption weights each path by $e^{-\mu_a L}$ in its total length $L$, which
truncates the subordinator and produces the screened radial structure
$(\mueff+1/r)e^{-\mueff r}/r^2$ carried by each source term of the
Farrell--Patterson--Wilson (1992) diffusion dipole; the finite-depth two-source
dipole is not derived here. Keeping the launch-depth factor that the tail
calculation discards continues the same kernel into a finite core of radius
$r_c=\sqrt{2\pi a}\,A_f\simeq1.7\,\ell^{*}$, set by two already-measured numbers
with no additional fitted parameter. A scale-dependent index $\aeff$ read off the
empirical characteristic function reparameterizes the similarity parameter
$\gamma$ over the experimentally accessible range, separating phase functions
matched in $g_1$ by $0.15$--$0.22$ near one transport mean free path; below half
a transport mean free path it separates phase functions matched in both $g_1$ and
$g_2$, by about three standard errors over eight seeds.
Results are Monte Carlo at $g\in\{0,0.5,0.8,0.9\}$ with up to $10^6$ walkers.
\end{abstract}

\section{Introduction}
\label{sec:intro}

The point-spread function of a turbid half space---the radial distribution of
where light re-emerges given a normally incident pencil beam---is the object a
spatially resolved reflectance measurement inverts for optical properties.
Farrell, Patterson and Wilson~\cite{farrell1992} gave the standard closed form:
a diffusion dipole, two image sources placed to satisfy the extrapolated
boundary condition. It fits Monte Carlo and phantom data down to about half a
transport mean free path and fails inside it, where the diffusion approximation
has no access to walkers that resolve after two or three collisions.

That near-source region, source--detector separations of order $\lstar$, is the
subdiffuse regime. Bevilacqua and Depeursinge~\cite{bevilacqua1999} showed by
Monte Carlo that reflectance there depends on the phase function beyond its first
moment, through a similarity parameter $\gamma=(1-g_2)/(1-g_1)$ built from the
first two Legendre moments $g_1,g_2$. Kienle, F\"orster and Hibst~\cite{kienle2001}
found that ignoring this dependence and fitting the standard diffusion solution
returns $\mu_s'$ and $\mu_a$ in error by up to a factor of two. Liemert and
Kienle~\cite{liemert2013} later solved the anisotropic transport equation for the
semi-infinite medium exactly, evaluated numerically.

We take a combinatorial route. The walk's return to the launch plane is a
first-passage problem, and in the diffusion limit its exit position is a
subordinated Brownian motion. This yields the far-field PSF as an isotropic $1$-stable
scaling limit, explains the $r^{-3}$ tail and the $\alpha=1$ index with no fitted
index,
and shows that absorption produces the same screened radial structure as the
Farrell dipole. It also gives a scale-resolved index $\aeff$ that reads the phase
function off the reflectance. The construction extends an earlier first-passage treatment of the Kubelka--Munk
problem~\cite{zeller2019} and the axial first-return machinery of
\cite{zeller2025motzkin,zeller2026cauchy}, which count returns by scattering
order but say nothing about where the walker lands. Three claims of
decreasing strength follow, and we keep them separate: the first-return law and
the Cauchy tail, which the simulations support directly; the structural link to
diffusion-dipole theory, which reproduces the radial form of a single screened
source but stops short of the finite-depth two-source construction; and $\aeff$ as a
phase-function statistic, which tracks $\gamma$ over the accessible range and
which we present as exploratory where it reaches past $\gamma$, below half a
transport mean free path.

\section{Model and Monte Carlo method}
\label{sec:model}

A walker starts at the origin moving along $+z$ into the half space $z>0$. It
takes exponentially distributed free paths with mean free path $\ell=1$ and
scattering coefficient $\mu_s=1$, scatters at each collision by the HG phase
function~\cite{henyey1941} of asymmetry $g$, and is recorded at its first
crossing of $z=0$. The
transport mean free path is $\lstar=\ell/(1-g)$ and the reduced scattering
coefficient $\mu_s'=1/\lstar$. Direction cosine on flight $i$ is $\mu_i$; the
depth increment is $\zeta_i=s_i\mu_i$. The record is $(x,y,0,\ns)$: the exit
point by exact linear interpolation along the crossing flight, and $\ns$ the
flight index at that crossing.

Throughout, $n$ counts flights, equivalently scattering events, and not
collisions of any other kind or units of path length: the walker takes flight $i$
from collision $i-1$ to collision $i$, the launch flight is $i=1$, and the flight
on which the walker crosses $z=0$ is counted. We write $\ns$ for the value of $n$
at that crossing when the distinction from a running index matters, and
$n_{\mathrm{cut}}$ for the ceiling. Walkers not returned by the ceiling are
censored; the ceiling is $10^4$ except where stated, and
Sec.~\ref{sec:alphaeff} reports what raising it changes.
The boundary is index-matched throughout, so the internal reflection parameter is
$A=1$. Table~\ref{tab:notation} collects the symbols.

\begin{table}[htbp]
  \centering
  \caption{Notation. Lengths are in units of the mean free path $\ell=1$.}
  \label{tab:notation}
  \begin{tabular}{ll}
    \toprule
    symbol & meaning \\
    \midrule
    $\ell$, $\mu_s$ & mean free path and scattering coefficient, both $1$ here \\
    $\lstar$, $\mu_s'$ & transport mean free path $\ell/(1-g)$ and its reciprocal \\
    $\mu_a$, $\mueff$ & absorption coefficient and screening rate $\sqrt{3\mu_a\mu_s'}$ \\
    $g$, $g_1$, $g_2$ & HG asymmetry; first and second Legendre moments \\
    $\gamma$ & similarity parameter $(1-g_2)/(1-g_1)$ \\
    $n$, $\ns$, $n_{\mathrm{cut}}$ & flight count; its value at exit; the ceiling \\
    $n_c$ & angular correlation length $-1/\ln g$, in flights \\
    $s_i$, $\mu_i$, $\zeta_i$ & length, direction cosine and depth increment of flight $i$ \\
    $L$ & total path length $\sum_i s_i$, including the crossing flight \\
    $Q(n)$, $f(n)$ & survival $P(T>n)$ and first-return density \\
    $A$, $A_f$ & survival amplitude, Eq.~\eqref{eq:qscale}; first-return amplitude \\
    $a$ & transverse variance per flight, $\tfrac23\ell\lstar$ \\
    $z_0$, $D_z$ & effective launch depth and vertical diffusion coefficient \\
    $r_c$ & core radius of the exit-density kernel, Eq.~\eqref{eq:core} \\
    $r$, $\rho$ & exit radius; source--detector separation \\
    $\alpha$, $\aeff$ & stable index; its scale-resolved estimate \\
    $k$, $\phi(k)$ & transverse wavenumber; empirical characteristic function \\
    \bottomrule
  \end{tabular}
\end{table}

Absorption is applied after the fact by importance reweighting rather than by a
separate run, so one non-absorbing ensemble serves every $\mu_a$. The exact weight
for a recorded trajectory of total path length $L=\sum_i s_i$ (including the
partial crossing flight) is $e^{-\mu_a L}$, and we store $L$ for every walker and
use it. A tempting shortcut replaces $L$ by the flight count, giving
$(1+\mu_a)^{-n}$ from $\mathbb{E}[e^{-\mu_a s}]=1/(1+\mu_a)$ per flight; this is
biased because conditioning on return at flight $n$ correlates the flight lengths.
We measured the bias directly: $\mathbb{E}[e^{-\mu_a L}\mid n]$ exceeds
$(1+\mu_a)^{-n}$ by a factor that grows without bound in $n$ (walks that return
late do so through many short flights, so their $L$ is far below $n\langle
s\rangle$), but those walks carry exponentially small weight, so the integrated
error in the reflectance is $0.1\%$ at $\mu_a=0.001$ and $1\%$ at $\mu_a=0.01$,
rising to a few percent locally at large radius. All results below use the exact
$e^{-\mu_a L}$ weight.

A $g=0$ run validates less of this code than it appears to, for a reason that is
a property of the theorem rather than of our implementation; Appendix
\ref{sec:repro} gives the case and lists the checks we rerun after any edit.

\section{The first-return law}
\label{sec:firstreturn}

For independent, identically distributed, symmetric continuous increments,
Sparre~Andersen~\cite{sparreandersen1954} fixes the survival and first-return
densities of the running sum,
\begin{equation}
Q(n)=\binom{2n}{n}4^{-n}\simeq(\pi n)^{-1/2},\qquad
f(n)=-\frac{dQ}{dn}\simeq\frac{1}{2\sqrt\pi}\,n^{-3/2}.
\label{eq:sparre}
\end{equation}
The survival $Q(n)$ is exact and discrete; the first-return density is the
difference $f(n)=Q(n-1)-Q(n)$, and we write it as $-dQ/dn$ because we use it
only through the continuum subordination that follows, where $n$ is treated as a
continuous time. The two agree to $O(1/n)$, below the resolution of every fit
reported here.

The exponent $-3/2$ is distribution-free: it survives any symmetric continuous
step law. What the theorem needs is independence, and HG scattering breaks it.
Successive directions are correlated,
\begin{equation}
\langle\mathbf u_i\cdot\mathbf u_{i+k}\rangle=g^{k},\qquad
n_c=-1/\ln g,
\label{eq:corr}
\end{equation}
so the depth increments $\zeta_i$ are correlated over $n_c$ flights. The exact
theorem therefore does not apply to the HG walk. What we claim, and test, is
weaker: the correlation is short-ranged, so a diffusion limit exists and the
walk inherits the $-3/2$ \emph{exponent} for $n\gg n_c$, while the finite-$n$ law
and the overall amplitude are process-dependent. The amplitude in particular is not given by the theorem for $g>0$; we fix it
empirically in Sec.~\ref{sec:survival}.

Stated as conditions, what the $-3/2$ exponent requires of the depth process
$\{\zeta_i\}$ is: stationarity; symmetry of the marginal increment law;
$\langle\zeta^2\rangle<\infty$; and summable correlations,
$\sum_k|\langle\zeta_i\zeta_{i+k}\rangle|<\infty$, which for the HG walk follows
from $\langle\mathbf u_i\cdot\mathbf u_{i+k}\rangle=g^k$ with $g<1$. Under these
the partial sums converge to Brownian motion under diffusive rescaling, and the
persistence exponent of the limit governs the tail of $f(n)$; the first two
conditions are what make the limit symmetric and the third and fourth what make
it Brownian rather than stable of lower index. The exponent is then $-3/2$
because Brownian first passage is, which is the same statement Sparre~Andersen
makes at the discrete level. Majumdar~\cite{majumdar2010} reviews the
distribution-free result and its scope, and Bray, Majumdar and
Schehr~\cite{bray2013} the persistence exponents of correlated processes, where
correlations that are \emph{not} summable do move the exponent. Sharp forward
scattering, $g\to1$, approaches that boundary at fixed $n$: $n_c$ diverges, and
the asymptotic regime moves out of reach of any finite ceiling rather than
ceasing to exist.

At $g=0$ the situation is cleaner than that, and worth stating because it is a
test. The depth increments $\zeta_i=s_i\mu_i$ with $s_i$ exponential and $\mu_i$
uniform on $[-1,1]$ are independent, symmetric and continuous, so Sparre~Andersen
applies to them exactly---amplitude included. What breaks the correspondence is
not the increments but the launch: the first flight is forced along $+z$, so the
walk starts at a random positive depth $s_1$ rather than at the origin. Sampling
the first direction from the same isotropic law as every other, and conditioning
on $\zeta_1>0$, must return $Q(n)\to2(\pi n)^{-1/2}$ with no fitted constant.
That is a sharper validation gate than any slope, and we list it with the other
checks in Appendix~\ref{sec:repro}.

Figure~\ref{fig:firstreturn} shows the measured $f(n)$ at four values of $g$.
The tails run parallel to the $n^{-3/2}$ reference; the small-$n$ shoulder grows
with $g$, extending to larger $n$ as $n_c$ grows. Fitting a power law through
that shoulder manufactures an exponent. Table~\ref{tab:slopes} makes the point
quantitative: at $g=0.8$ a window opened at $n>5$ reads $-1.325$, off the
asymptote by $0.17$; the exponent recovers to $-1.498$ only once the window
clears the crossover.

\begin{figure}[htbp]
  \centering
  \includegraphics[width=0.62\textwidth]{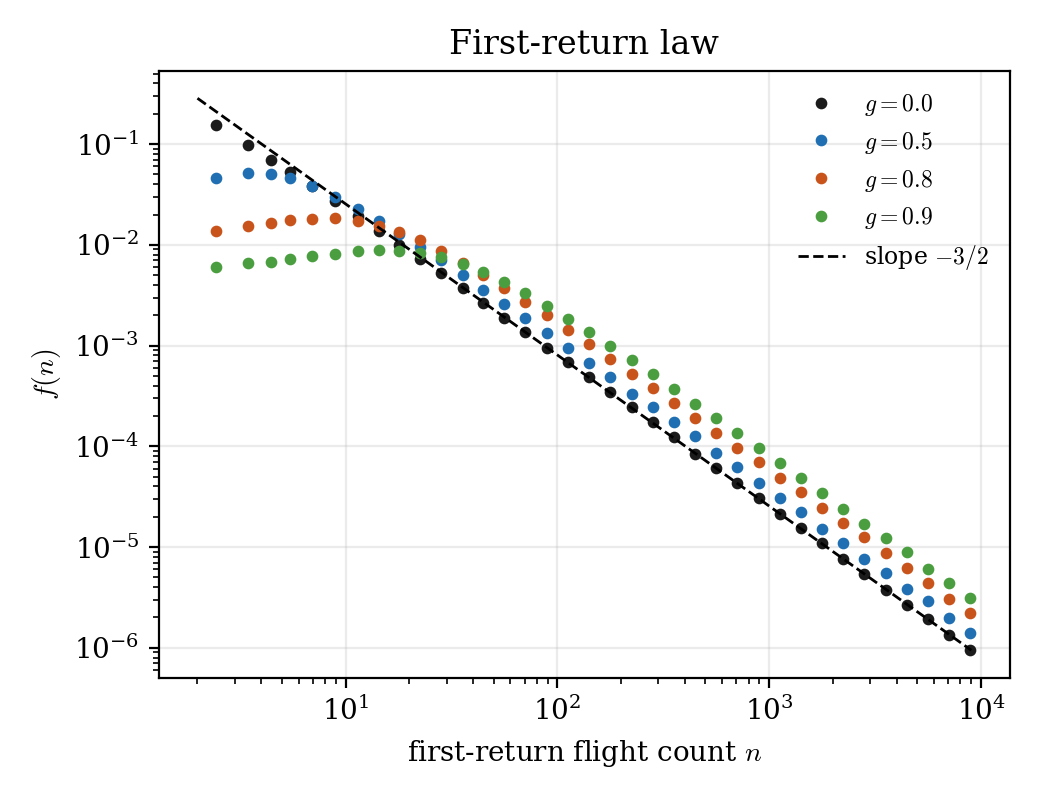}
  \caption{First-return density $f(n)$ for $g\in\{0,0.5,0.8,0.9\}$, from
    $10^6$ ($g=0,0.8$) and $5\times10^5$ ($g=0.5,0.9$) walkers. The dashed line
    has slope $-3/2$, anchored to the $g=0$ tail. Correlation
    (Eq.~\ref{eq:corr}) produces a shoulder at $n\lesssim n_c$ that widens with
    $g$; the tails share the distribution-free exponent.}
  \label{fig:firstreturn}
\end{figure}

\begin{table}[htbp]
  \centering
  \caption{First-return slope by fit window. The asymptote is $-3/2$; fits
    opened below the crossover $n_c=-1/\ln g$ read high.}
  \label{tab:slopes}
  \begin{tabular}{lcc}
    \toprule
    window & $g=0$ & $g=0.8$ \\
    \midrule
    $n>5$    & $-1.485$ & $-1.325$ \\
    $n>50$   & $-1.502$ & $-1.473$ \\
    $n>300$  & $-1.506$ & $-1.491$ \\
    $n>1000$ & $-1.507$ & $-1.498$ \\
    \bottomrule
  \end{tabular}
\end{table}

\section{The survival amplitude scales as $\sqrt{\lstar}$}
\label{sec:survival}

\begin{figure}[htbp]
  \centering
  \includegraphics[width=0.92\textwidth]{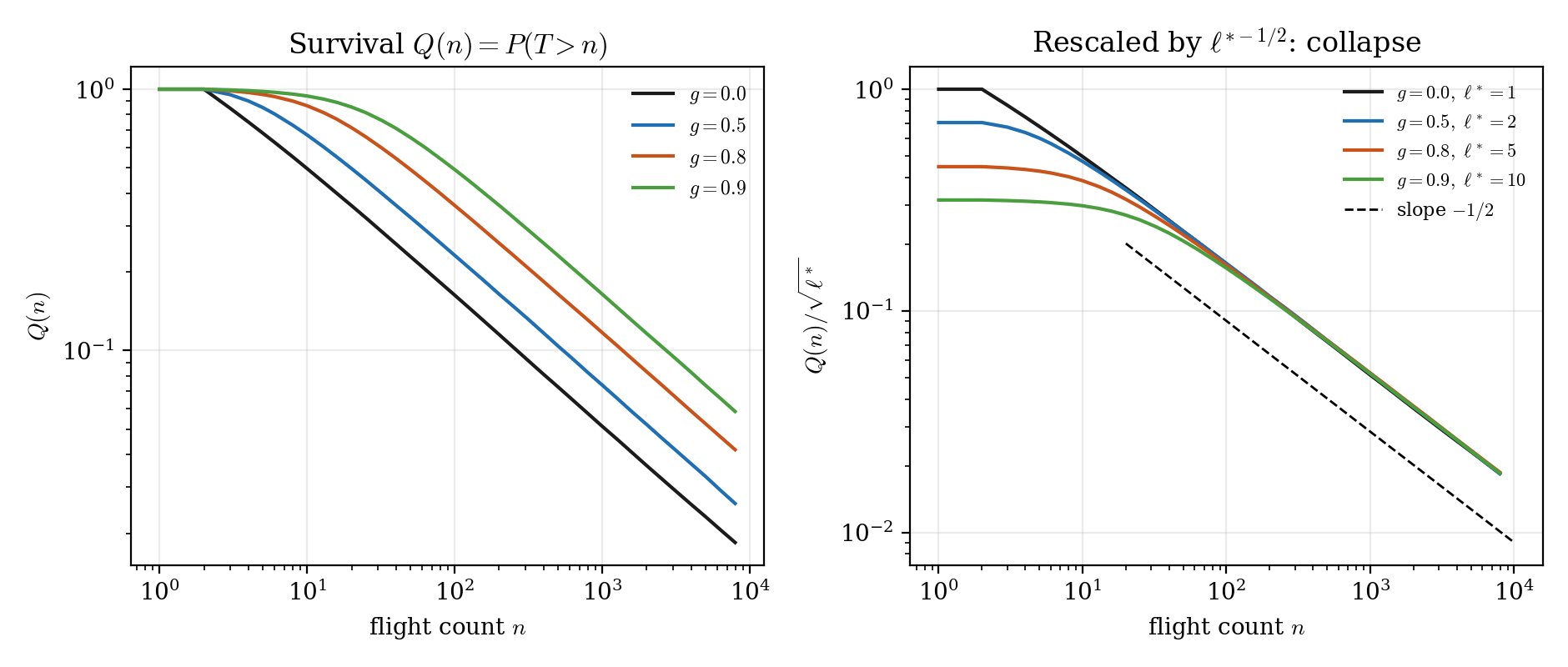}
  \caption{Left: survival $Q(n)=P(T>n)$ for the four $g$, separated in the tail.
    Right: rescaled by $1/\sqrt{\lstar}$ the curves collapse onto a single
    $n^{-1/2}$ law (dashed), confirming Eq.~\eqref{eq:qscale}. The low-$n$
    plateaus are the pre-crossover regime.}
  \label{fig:survival}
\end{figure}

Correlation renormalizes the diffusion constant, $\langle R^2\rangle_n=2\ell\lstar
n$, and rescales the effective launch depth by the same $\lstar$. In the Brownian
limit $Q(n)=z_0/\sqrt{\pi D n}$, and the two factors combine to
\begin{equation}
Q(n)\simeq A\,\sqrt{\lstar/n}.
\label{eq:qscale}
\end{equation}
Here $A$ is a single amplitude, not fixed by the theorem; we set it on the $g=0$
run and hold it across $g$, so Eq.~\eqref{eq:qscale} is a one-parameter prediction
for all other $g$. The prediction is a data collapse: dividing the measured
survival by $\sqrt{\lstar}$ should map every $g$ onto one curve of slope
$-\tfrac12$.
Figure~\ref{fig:survival} shows it does. The raw survival curves separate by an
order of magnitude in the tail; rescaled by $1/\sqrt{\lstar}$ they lie on top of
one another for $n$ past the crossover, along the $n^{-1/2}$ guide.

Anchoring $A$ on $g=0$ and predicting the non-absorbing return fraction at
$n=10^4$ tests Eq.~\eqref{eq:qscale} against absolute numbers. The prediction
$Q\propto\sqrt{\lstar}$ holds to Monte Carlo accuracy: $2.331\%$ predicted
against $2.312\%$ observed at $g=0.5$, $3.685\%$ against $3.721\%$ at $g=0.8$,
$5.212\%$ against $5.203\%$ at $g=0.9$, each within the $\sim1\%$ sampling error.
These numbers correspond to $A\simeq1.64$.

That value is about $2.9$ times the Sparre~Andersen amplitude $1/\sqrt\pi$, and
the launch accounts for the factor. Conditioning the first increment to be
positive contributes $2$; the rest is the first flight being longer than a
typical positive increment, $\langle s\rangle=1$ against
$\langle s\mu\mid\mu>0\rangle=\tfrac12$, softened near the boundary because the
Brownian relation $Q\propto z_0$ needs $z_0$ large against a step. So $A$ is a
launch constant, not a free parameter of the walk, and Sec.~\ref{sec:core} shows
it is the only constant the far-field kernel needs.

\section{The exit density, its core, and the moments that do not exist}
\label{sec:exitdensity}

\begin{figure}[htbp]
  \centering
  \includegraphics[width=0.92\textwidth]{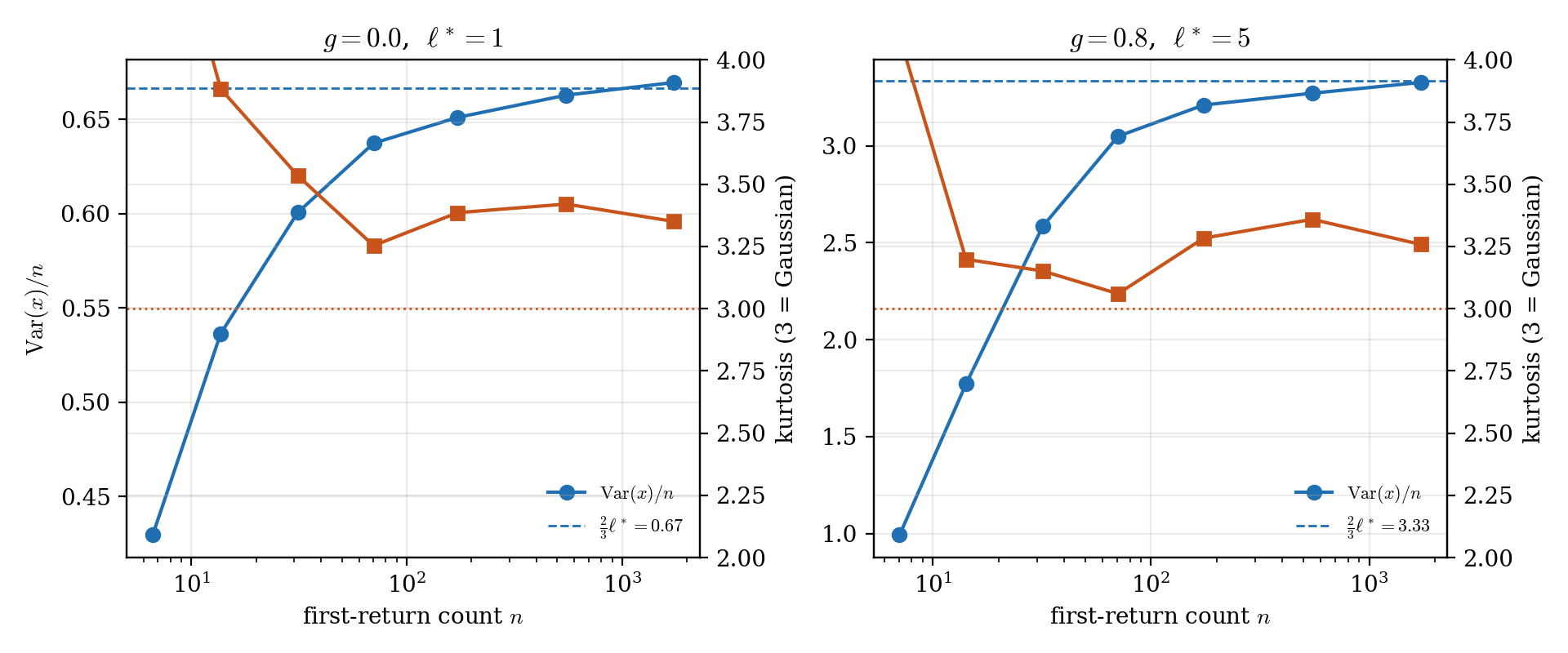}
  \caption{Test of the conditional Gaussianity assumed by the subordination.
    Return-conditioned transverse variance per flight $\mathrm{Var}(x)/n$ (blue,
    left axis) approaches $\tfrac23\ell\lstar$ (dashed) and kurtosis (orange,
    right axis) approaches $3$ (dotted) as $n$ grows, both accurate to a few
    percent for $n\gtrsim100$. The small-$n$ departure is the correlated
    ballistic transient.}
  \label{fig:gauss}
\end{figure}

The subordination rests on one assumption: that the lateral displacement,
conditioned on first return at flight $n$, is Gaussian with variance growing
linearly in $n$. We test it directly (Fig.~\ref{fig:gauss}). Binning returned
walkers by $n$ and pooling the two Cartesian components, the conditional variance
per flight $\mathrm{Var}(x)/n$ rises from below and settles onto
$\tfrac23\ell\lstar$---$0.670$ against $0.667$ at $g=0$, $3.33$ against $3.33$ at
$g=0.8$---while the kurtosis falls from above toward the Gaussian value $3$. The
variance reaches within $4\%$ of $\tfrac23\ell\lstar$ by $n\approx200$ (from
$36\%$ low at $n\approx7$ for $g=0$, more for $g=0.8$), and the excess kurtosis,
near $1$ at
$n\approx7$, drops to a few tenths for $n\gtrsim15$, with the residual scatter
there set by the kurtosis estimator rather than by non-Gaussianity. The Gaussian conditioning is thus a large-$n$ statement, accurate exactly where
the $-3/2$ first-return exponent also holds and failing together with it in the
ballistic near field, where the walk has not yet decorrelated.

One point about what this test buys. Bochner subordination in its usual
statement needs the subordinator independent of the Brownian motion, and the
first-return time here is not independent of the transverse walk: both are built
from the same flight lengths $s_i$. Independence is more than the construction
needs. It is enough that the law of $(x,y)$ given first return at flight $n$
depend on $n$ only through its variance $an$, and that is what
Fig.~\ref{fig:gauss} measures. We state the assumption in that form and test it
in that form.

\begin{figure}[htbp]
  \centering
  \includegraphics[width=0.62\textwidth]{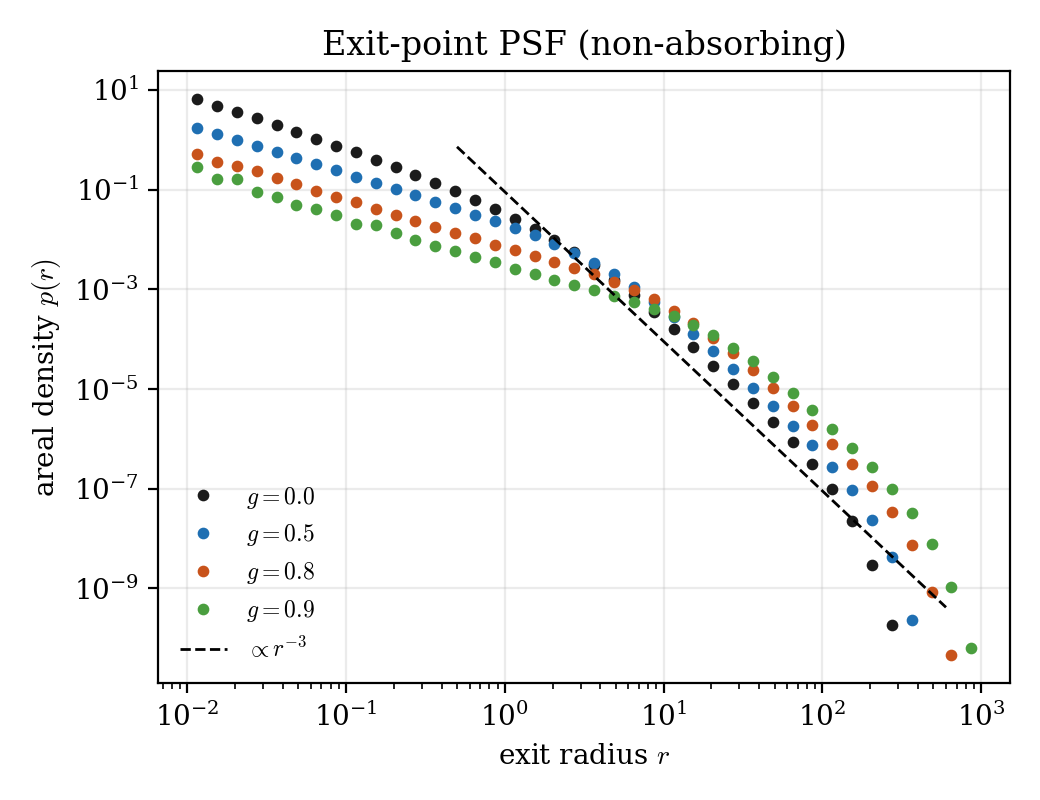}
  \caption{Non-absorbing exit-point density (the PSF) versus radius, as areal
    density: counts per logarithmic radial bin ($40$ log-uniform bins from
    $r=0.01$ to $1000$, eight per decade)
    divided by annulus area $\pi(r_{i+1}^2-r_i^2)$ and by the walker count. The
    dashed guide has slope $-3$, the Cauchy / Poisson-kernel tail of
    Eq.~\eqref{eq:pr}. The high-$r$ rollover is the finite-$\ns$ cutoff, not a
    change in the law; Fig.~\ref{fig:ceiling}(b) shows it receding as the ceiling
    is raised.}
  \label{fig:psf}
\end{figure}

With that assumption, marginalize the transverse Gaussian over the first-return
law. The continuum (Brownian) model for the vertical first-passage, with launch
depth $z_0$ and vertical diffusion $D_z$, is the L\'evy--Smirnov density
\begin{equation}
f(n)=\frac{z_0}{\sqrt{4\pi D_z}}\,n^{-3/2}\,
      \exp\!\Big(\!-\frac{z_0^2}{4 D_z n}\Big).
\label{eq:levy}
\end{equation}
Its large-$n$ limit is $A_f\,n^{-3/2}$ with $A_f=z_0/\sqrt{4\pi D_z}$. This
shares the exponent of Eq.~\eqref{eq:sparre}, but not its coefficient: $A_f$ is
the process-dependent amplitude of Sec.~\ref{sec:survival}, related to the
survival amplitude by $A_f=\tfrac12 A\sqrt{\lstar}$ (since $f=-dQ/dn$), and equal
to $A_f\simeq0.82$ on the $g=0$ run---not the simple-lattice-walk value
$1/(2\sqrt\pi)\simeq0.28$. Reading Eq.~\eqref{eq:levy} the other way, matching
$A_f$ calibrates the effective combination $z_0/\sqrt{D_z}=\sqrt{4\pi}\,A_f$;
Eq.~\eqref{eq:levy} is thus a Brownian model whose effective launch depth is set
by the measured amplitude, not assigned. Integrating the \emph{full} density
\eqref{eq:levy}
against the Gaussian is the classical subordination integral and returns the
finite-core half-space Poisson kernel $p(r)=(z_0/2\pi)(z_0^2+r^2)^{-3/2}$, the
isotropic $2$-D Cauchy law. Integrating only its large-$n$ tail, on the other
hand, drops the launch-depth factor and keeps only the tail of that kernel,
\begin{equation}
p(r)\;\propto\;r^{-3}\!\int_0^\infty t^{1/2}e^{-t}\,dt\;\propto\;r^{-3}
\qquad (r\gg z_0),
\label{eq:pr}
\end{equation}
with $t=r^2/2\sigma_n^2$. Equation~\eqref{eq:pr} is the far-field tail, valid for
$r$ beyond the launch depth $z_0\sim\lstar$; it is not the normalized Cauchy
density and is singular as $r\to0$, where the finite core of the full kernel
takes over. The simulations resolve the tail, not the core:
Figure~\ref{fig:psf} shows the measured areal density following $r^{-3}$ over the
intermediate decades at every $g$, rolling over only where the $\ns\le10^4$
ceiling truncates the longest walks.

Quoting a slope means quoting the window it was fitted over, because the core
truncates that window from below and the ceiling from above. Least squares
outside three core radii, $r>3r_c\simeq5\lstar$ with $r_c$ from
Eq.~\eqref{eq:rc} below, and inside the rollover gives $-3.07$, $-2.98$, $-2.90$
and $-2.87$ at $g=0$, $0.5$, $0.8$ and $0.9$. Every slope quoted in this paper
uses that window. A window fixed in absolute units reads lower and drifts with
$g$: $1<r<100$ gives $-2.66$ at $g=0$ and $-1.85$ at $g=0.8$, since $r_c$ scales
as $\lstar$ and swallows most of that window by $g=0.8$.
Section~\ref{sec:core} continues the same
Brownian model inward to a finite core, using the same two measured numbers.

\subsection{The core scale is fixed by the same two measurements}
\label{sec:core}

Dropping the launch-depth factor was a choice. Carrying the
\emph{full} density~\eqref{eq:levy} through the Gaussian marginalization costs one
Gamma integral, $\int_0^\infty n^{-5/2}e^{-B/n}\,dn=\Gamma(\tfrac32)B^{-3/2}$
with $B=z_0^2/4D_z+r^2/2a$, and gives the normalized half-space Poisson kernel
outright,
\begin{equation}
p(r)=\frac{r_c}{2\pi}\,\big(r_c^2+r^2\big)^{-3/2},\qquad
r_c=\sqrt{2\pi a}\;A_f .
\label{eq:core}
\end{equation}
Both inputs are already measured: $a=\tfrac23\ell\lstar$ is the transverse
variance per flight of Fig.~\ref{fig:gauss}, and $A_f$ is the first-return
amplitude of Sec.~\ref{sec:survival}. The step that makes this work is
$z_0^2/4D_z=\pi A_f^2$, which follows from $A_f=z_0/\sqrt{4\pi D_z}$: the
amplitude we fitted in Sec.~\ref{sec:survival} \emph{is} the core scale in
disguise, so no effective launch depth has to be assigned separately. The
anisotropy of the correlated walk drops out for the same reason: $A_f$ carries
the vertical statistics and $a$ carries the transverse, so the ratio
$D_z/D_{xy}$ is never needed as a separate input.

Both factors carry their $\lstar$ scaling with them, $a\propto\lstar$ and
$A_f\propto\sqrt{\lstar}$, leaving
\begin{equation}
r_c=A\sqrt{\pi/3}\;\lstar\simeq1.68\,\lstar
\label{eq:rc}
\end{equation}
in units of $\ell=1$, with $A\simeq1.64$ from Sec.~\ref{sec:survival}. The core
radius carries no additional fitted parameter; it is the launch constant $A$
again, wearing different units.

What Eq.~\eqref{eq:core} is, and is not, deserves a sentence, since the core is
the part of the kernel our derivation reaches by continuation rather than by
limit. The far-field tail follows from the subordination and the two theorems
behind Eq.~\eqref{eq:alpha}, and holds for any walk meeting the conditions of
Sec.~\ref{sec:firstreturn}. Equation~\eqref{eq:core} instead evaluates the
Brownian model~\eqref{eq:levy} at radii where the walk it models has not
decorrelated, and where Fig.~\ref{fig:gauss} shows the conditional Gaussianity
failing. It predicts a specific $r_c$ from measured numbers, which makes it
testable, and we test it below; it does not inherit the universality of the tail.

Two independent readings of the simulations agree with Eq.~\eqref{eq:rc} to about
$10\%$. One of them is the cutoff dependence of $\langle r\rangle$, which comes
next; the other is the scale parameter of the characteristic function, which
waits for Sec.~\ref{sec:alphaeff}.

\subsection{The moments that do not exist}

An $r^{-3}$ areal density means a radial density $P(r)=2\pi r\,p(r)\propto
r^{-2}$, and its low moments diverge. Truncating the walk at $n_{\mathrm{cut}}$
truncates the radius at $r_{\max}\sim\sqrt{\ell\lstar n_{\mathrm{cut}}}$, and
Eq.~\eqref{eq:core} fixes the prefactors as well as the scalings,
\begin{equation}
\langle r\rangle\simeq r_c\big[\ln(2r_{\max}/r_c)-1\big],\qquad
\langle r^2\rangle\simeq r_c\,r_{\max},
\label{eq:moments}
\end{equation}
neither convergent. Both grow with the cutoff, so a reported $\langle
r^2\rangle$ measures $n_{\mathrm{cut}}$ as much as the medium.

The per-decade increment of $\langle r\rangle$ is $\tfrac12 r_c\ln10\simeq1.15
\,r_c$, independent of where the core is cut off, so it measures $r_c$ directly.
Per decade the increments settle at $1.77$ ($g=0$) and $8.62$ ($g=0.8$), giving
$r_c=1.54$ and $1.50\,\lstar$; the ratio $4.9$ against the predicted $\lstar=5$
is the same statement read as a scaling. The $\langle r^2\rangle$ ratios approach
$3.29$ and $3.44$ against $\sqrt{10}=3.16$. The median, by contrast, moves only
$5\%$ over three decades of cutoff. Report quantiles, or fit the CDF of
Eq.~\eqref{eq:core} directly.

Three routes to the core scale therefore close on each other:
Eq.~\eqref{eq:rc} predicts $1.68\,\lstar$ from the survival amplitude and the
conditional variance; the $\langle r\rangle$ increments give $1.50$--$1.54\,
\lstar$; and the scale parameter of the empirical characteristic function
(Sec.~\ref{sec:alphaeff}) gives $1.3$--$1.5\,\lstar$. The spread is what one
should expect from a kernel whose measured index on the plateau is $0.88$--$0.96$
rather than exactly $1$, and it is a sharper consistency check on the
subordination than the tail slope, which only tests an exponent.

The divergence is not a sampling artifact to be managed. An infinite variance is
the defining property of a stable law of index below $2$, and for the Cauchy case
even the mean fails to converge. The $r^{-3}$ tail and the non-existent second
moment are the same fact, and Sec.~\ref{sec:tempered} names the law behind both.

\section{The far-field PSF as a stable law, and its link to Farrell's dipole}
\label{sec:tempered}

\paragraph{The stable index is forced, not fitted.}
The first-return count has the asymptotic density $f(n)\propto n^{-3/2}$, the
tail of the one-sided $\tfrac12$-stable (L\'evy--Smirnov) law. The lateral
displacement conditioned on $n$ is Gaussian for $n$ past the transient
(Sec.~\ref{sec:exitdensity}, Fig.~\ref{fig:gauss}). Bochner subordination of a
Brownian motion by a one-sided $\alpha'$-stable subordinator produces a symmetric
$2\alpha'$-stable law, so in the diffusion limit
\begin{equation}
\alpha=2\alpha'=1.
\label{eq:alpha}
\end{equation}
The two factors are independent theorems. The $2$ is the central limit theorem:
the transverse coordinate at fixed $n$ is Gaussian. The $\tfrac12$ is
Sparre~Andersen: the first-return exponent is distribution-free. Neither is
adjustable, so no phase function with short-range angular correlation and finite
step variance---one admitting the diffusion limit these two theorems assume---can
move the far-field index off $1$. Only the scale parameter $c\propto\lstar$
depends on $g$; every such walk, isotropic to sharply forward, collapses onto the
same $1$-stable shape, rescaled.

The scaling limit is isotropic $1$-stable, the two-dimensional Cauchy law, whose
tail is the Poisson kernel of Sec.~\ref{sec:exitdensity}---the harmonic measure
of a Brownian motion crossing a plane at depth $z_0$, a classical result that the
subordination here recovers from the scattering microdynamics. The $r^{-3}$ tail
read off the data and the $\alpha=1$ scaling law derived from subordination are
one statement reached two ways. The statement is a far-field one---it fixes the tail exponent and the index---but
the core is not out of reach either, and Sec.~\ref{sec:core} recovers it from the
full first-passage density~\eqref{eq:levy}.

\paragraph{Absorption gives the screened radial structure of the dipole.}
Absorption weights each path by $e^{-\mu_a L}$ in its total length $L$
(Sec.~\ref{sec:model}). To keep the subordinator analytic we make one
approximation: replace the path length by the count time, $L\to n\langle
s\rangle=n\ell$, so the weight becomes $e^{-\mu_a\ell n}$. The factor $\ell$ is
invisible in our units but carries the dimensions, and restoring it is what makes
the screening rate below come out right rather than come out lucky. Carrying that count-time weight
through the marginalization of the bare tail \eqref{eq:pr},
\begin{equation}
p(r)\propto\int_0^\infty n^{-5/2}\exp\!\Big(-\mu_a n-\frac{r^2}{2an}\Big)dn
=2\Big(\frac{r^2}{2a\mu_a}\Big)^{-3/4}K_{3/2}(\lambda r),
\label{eq:tempered}
\end{equation}
with $a=\tfrac23\ell\lstar$ and $\lambda=\sqrt{2\mu_a\ell/a}$. Using the closed
form $K_{3/2}(x)=\sqrt{\pi/2x}\,e^{-x}(1+1/x)$,
\begin{equation}
p(r)\propto\Big(\lambda+\frac1r\Big)\frac{e^{-\lambda r}}{r^2},\qquad
\lambda=\sqrt{3\mu_a\mu_s'}=\mueff.
\label{eq:farrellform}
\end{equation}
The screening rate is the textbook one, and the reason is worth one line: the
transverse variance per unit path length is $a/2\ell=\tfrac13\lstar$, which is
the diffusion coefficient $D=1/(3\mu_s')$, so $\lambda=\sqrt{\mu_a/D}$ by
construction. The subordination does not produce a new screening length; it
produces the diffusion one, from the walk statistics rather than from the
diffusion equation.

Equation~\eqref{eq:farrellform} is the low-absorption (count-time) form of the
screened density. Writing the replacement $L\to n\ell$ as a cumulant truncation
shows what it costs. Conditioned on return at flight $n$, the exact weight is
\begin{equation}
\mathbb{E}\!\left[e^{-\mu_a L}\mid n\right]
  =\exp\!\Big[-\mu_a\langle L\mid n\rangle
    +\tfrac12\mu_a^2\,\mathrm{Var}(L\mid n)-\cdots\Big],
\label{eq:cumulant}
\end{equation}
so the count-time form makes two separate errors: it replaces
$\langle L\mid n\rangle$ by $n\ell$, and it discards the variance and the higher
cumulants. Both $\langle L\mid n\rangle$ and $\mathrm{Var}(L\mid n)$ belong to
the non-absorbing ensemble and are independent of $\mu_a$; what controls the
error is the exponent, $\mu_a\big(n\ell-\langle L\mid n\rangle\big)$ at first
order and $\tfrac12\mu_a^2\mathrm{Var}(L\mid n)$ at second. The first-order term
is the bias measured in Sec.~\ref{sec:model}: $\langle L\mid n\rangle$ falls
below $n\ell$ at large $n$, since walks that return late do so through many short
flights. Those $n$ carry exponentially small weight once $\mu_a>0$, so the
integrated error stays where Table~\ref{tab:farrell} puts it, $0.3\%$ at
$\mu_a=0.001$ and a few percent at $\mu_a=0.01$. The screening rate
$\mueff=\sqrt{3\mu_a\mu_s'}$ is the leading term of the tempering rate, and the
$O(\mu_a)$ correction to it comes from the same two cumulants; the full
path-length subordinator we leave open.

This is the screened radial structure of a Farrell--Patterson--Wilson source
term~\cite{farrell1992}, with $\mueff$ as the screening rate. It is not yet the
dipole. Equation~\eqref{eq:tempered} starts from the bare $n^{-3/2}$ tail, i.e.\
from zero source depth, so it yields a single screened source at the origin.
Farrell's model places two finite-depth sources, a real source at $z_0$ and an
image at $z_0+2z_b$, at distances $r_i=\sqrt{\rho^2+z_i^2}$ set by an
outward-flux boundary condition. Recovering both terms with their depths and
coefficients requires retaining the launch-depth factor of Eq.~\eqref{eq:levy}
through the marginalization and distinguishing fluence from outward flux, which
we have not carried out. The claim we defend is the shared radial form
$(\mueff+1/r)e^{-\mueff r}/r^2$, not a term-by-term identity of the full dipole.

Three quantities stay distinct throughout, and the comparison below is only
meaningful if they are kept so. Our $p(r)$ is an exit-point density: the
probability per unit area that a walker makes its first crossing of $z=0$ at
radius $r$, normalized by the launched walker count. Farrell's $R(\rho)$ is an
outward flux at the boundary, obtained from the diffusion fluence
$\Phi(\rho,z)$ by Fick's law at $z=0$. For an index-matched boundary and in the
diffusion limit the two are proportional, which is what makes
Table~\ref{tab:farrell} a ratio worth quoting, and inside $\lstar$ neither the
proportionality nor the fluence that underlies it is reliable.

\paragraph{Numerical check.}
Weighting the non-absorbing $g=0.8$ sample by the exact $e^{-\mu_a L}$ and
comparing against the Farrell dipole gives three regimes
(Table~\ref{tab:farrell}): agreement within a few percent beyond $\sim8\lstar$; a
$10$--$15\%$ shortfall between $\lstar$ and $4\lstar$; and a Monte Carlo excess
reaching $1.7$ inside $\lstar$, where diffusion theory has no access to walkers
returning after two or three collisions, and the far-field stable law, built on
the diffusive conditional variance, does not either. The count-weight shortcut
$(1+\mu_a)^{-n}$ (last column) tracks the exact weighting to $0.3\%$ at
$\mu_a=0.001$ but departs by up to $6\%$ at $\mu_a=0.01$ and large $\rho$, where
long paths dominate; the qualitative three-regime picture is unchanged.

\begin{table}[htbp]
  \centering
  \caption{Ratio of the Monte Carlo PSF to the Farrell dipole at $g=0.8$
    ($\lstar=5$), with the exact path-length weight $e^{-\mu_a L}$. The last
    column repeats $\mu_a=0.01$ with the biased count weight $(1+\mu_a)^{-n}$ for
    comparison. Agreement is good beyond a few $\lstar$; both models miss the
    near-source excess.}
  \label{tab:farrell}
  \begin{tabular}{cccc}
    \toprule
    $\rho$ & $e^{-\mu_a L}$, $\mu_a=0.001$ & $e^{-\mu_a L}$, $\mu_a=0.01$
           & $(1+\mu_a)^{-n}$, $\mu_a=0.01$ \\
    \midrule
    1.1 & 1.736 & 1.624 & 1.592 \\
    2.3 & 1.077 & 0.992 & 0.974 \\
    4.8 & 0.874 & 0.810 & 0.797 \\
    9.8 & 0.897 & 0.847 & 0.839 \\
    20  & 0.946 & 0.911 & 0.914 \\
    41  & 0.985 & 0.986 & 1.009 \\
    84  & 1.006 & 1.145 & 1.207 \\
    \bottomrule
  \end{tabular}
\end{table}

\section{A scale-resolved index that reads the phase function}
\label{sec:alphaeff}

The far-field index is $1$, and it moves away from $1$ at finite scale. To read
it off the data we use one estimator throughout, specified here in full, since
$\aeff$ is a local slope and its value depends on the grid it is taken on.

\begin{figure}[htbp]
  \centering
  \includegraphics[width=0.6\textwidth]{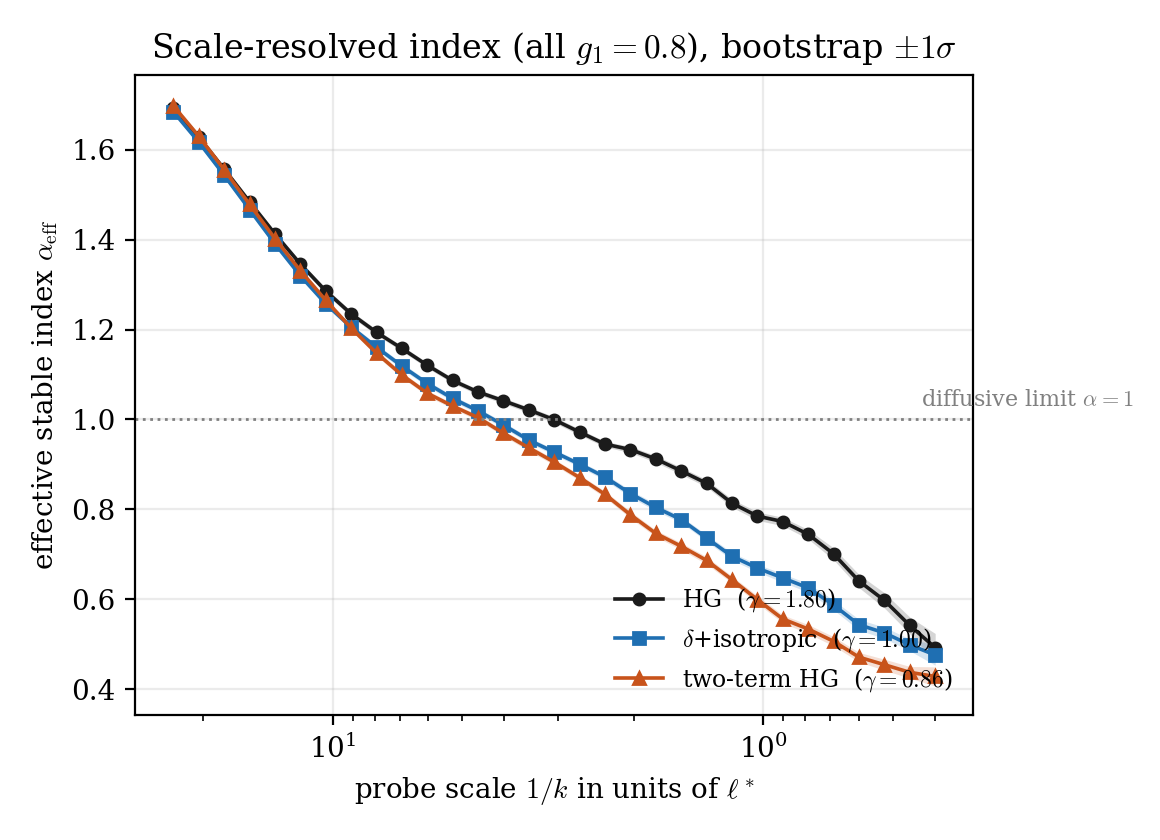}
  \caption{Effective stable index $\aeff$ versus probe scale $1/k$ (in units of
    $\lstar$) for three phase functions with matched $g_1=0.8$, from
    $4\times10^5$ walkers each. Shaded bands are $\pm1\sigma$ from
    $300$-resample walker bootstrap; they are narrow ($\sigma\lesssim0.008$ above
    $\lstar$, reaching $0.03$ at $0.4\lstar$). Curves separate below $\sim2\lstar$
    in order of $\gamma$, by up to $0.15$--$0.22$ near $\lstar$---twenty times the
    bootstrap width. The common rise above $1$ at the largest scales is the
    finite-$\ns$ cutoff lightening the tail, not a physical index above unity
    (Fig.~\ref{fig:ceiling}a); the intrinsic far-field value is $1$.}
  \label{fig:alphaeff}
\end{figure}

\begin{figure}[htbp]
  \centering
  \includegraphics[width=0.95\textwidth]{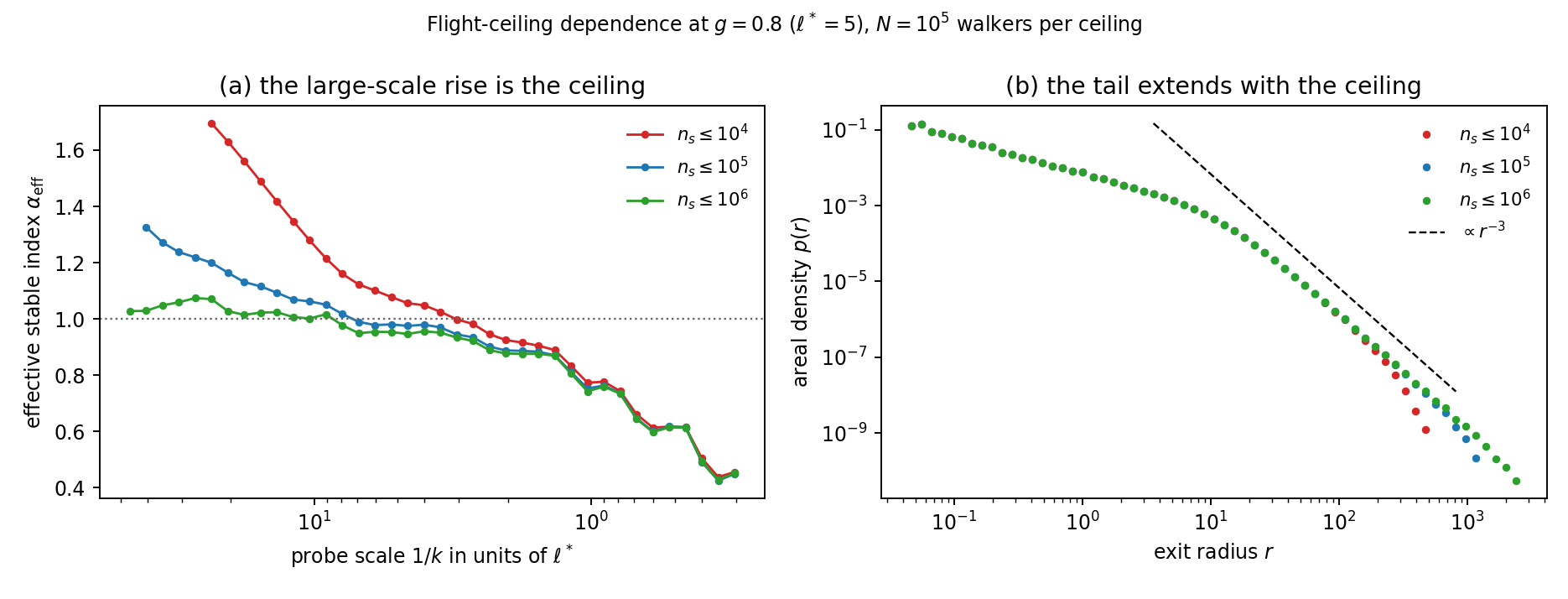}
  \caption{Flight-ceiling dependence at $g=0.8$ ($\lstar=5$), $10^5$ walkers per
    ceiling. (a) The rise of $\aeff$ above $1$ at large probe scale falls from
    $1.70$ to $1.33$ to $1.07$ as the ceiling goes $10^4\to10^5\to10^6$: it is
    the cutoff, not a physical index above unity. Below $\lstar$ the three curves
    lie on top of one another. (b) The range over which the areal density holds
    the $r^{-3}$ slope extends from $22\lstar$ to $79\lstar$ to $163\lstar$; the
    rollover in Fig.~\ref{fig:psf} moves with the ceiling and is not a change in
    the law.}
  \label{fig:ceiling}
\end{figure}

Form the empirical characteristic function of a Cartesian marginal,
$\phi(k)=\langle\cos kx\rangle$, averaged over all returned uncensored walkers
with the $x$ and $y$ marginals pooled. Evaluate it on a uniform logarithmic grid
running from $k=10^{-2.6}$ to $10^{-0.3}$ in units of $\ell^{-1}$, extended by
two further points beyond $10^{-0.3}$ at the same spacing. The plotted points of
Figs.~\ref{fig:alphaeff} and~\ref{fig:ceiling}(a) are that grid, so its spacing can
be read off them. A symmetric stable
law has $\phi(k)=\exp[-(ck)^\alpha]$, so the local slope of $\ln(-\ln\phi)$
against $\ln k$ gives an effective index $\aeff$ at probe scale $1/k$. We take
that slope by centred difference at every interior grid point, and report it only
where $0.01<\phi<0.97$; outside that window $-\ln\phi$ is dominated by sampling
noise at large $k$ and by its own vanishing at small $k$. Neighbouring $k$
estimates are correlated, so the curve is smoother than its point count suggests,
and we attach no independent error bar to individual points.

Two scales are quoted throughout, both named in units of $\lstar$: a plateau
value, the mean of $\aeff$ over $2$--$4\,\lstar$, and a near-source value at
$0.4\,\lstar$. The near-source scale is the smallest the estimator reaches: at
$\lstar=5$ it is the last interior point of the extended grid that falls inside
the $\phi$ window, so the grid and the window place it; the recipe above fixes
it without reference to any phase-function comparison. These two scales are the
only ones we quote.
The grid extension is what makes it interior; on the unextended grid it is the
endpoint and picks up a one-sided difference, which more than doubles its
seed-to-seed scatter, from $0.011$ to $0.026$. The extension costs no extra
walkers.

A pure $2$-D Cauchy has $\phi(k)=e^{-ck}$ and $\aeff=1$ at every scale, so the
measured departures are the PSF failing to be exactly Cauchy. Two biases act in
opposite directions. Below $\lstar$, single and double scattering sharpen the
core and pull $\aeff$ under $1$. The finite-$\ns$ ceiling truncates the longest
walks, removing the far tail and lightening the distribution toward Gaussian,
which pushes $\aeff$ up. Raising the ceiling shows this directly
(Fig.~\ref{fig:ceiling}): at $g=0.8$ the largest-scale index falls from $1.70$ at
$\ns\le10^4$ to $1.33$ at $10^5$ and $1.07$ at $10^6$, heading to the value
Eq.~\eqref{eq:alpha} requires.

The bias is not confined to the largest probed scales. It reaches into the
intermediate plateau, lifting $\aeff$ over $2$--$4\,\lstar$ by
$0.024$ at $g=0$ and by $0.078$ at $g=0.9$. Quoted at a $10^5$ ceiling, where the
residual shift to $10^6$ is under $0.01$, the plateau index over $2$--$4\,\lstar$
reads $0.882$, $0.907$, $0.933$ and $0.956$ at $g=0$, $0.5$, $0.8$ and $0.9$,
with scale parameter $c\simeq1.3$--$1.5\,\lstar$, consistent with the core radius
$r_c$ of Eq.~\eqref{eq:rc}. The plateau index rises monotonically with $g$ for a
reason that has nothing to do with the phase function: a fixed window in units of
$\lstar$ costs $n\simeq R^2/2\ell\lstar$ flights to reach, so at $R=3\lstar$ the
walk has taken about $5$ flights at $g=0$ and about $45$ at $g=0.9$. The same
probe scale samples a walk ten times further into its diffusion limit, and the
index sits correspondingly closer to $1$. The near-source end is insensitive to
the ceiling---at $0.4\,\lstar$ the three ceilings give $0.504$,
$0.491$ and $0.493$---so the separations reported below are not cutoff artifacts.

Because $\aeff$ is scale-resolved, the scale at which it is quoted has to be
named, and named in units of $\lstar$. A window fixed in $k$ rather than in
$\lstar$ covers $1.5$--$400\,\lstar$ at $g=0$ but $0.15$--$40\,\lstar$ at
$g=0.9$, so it samples the two walks in different regimes and reverses the
apparent trend with $g$. The plateau values above are all taken over the same
$2$--$4\,\lstar$ window.

The index separates phase functions matched in $g_1$. We take three with common
$g_1=0.8$, hence common $\lstar=5$: HG ($g_2=0.640$, $\gamma=1.80$), a
$\delta$-plus-isotropic mixture ($g_2=0.800$, $\gamma=1.00$), and a two-term HG
($g_2=0.828$, $\gamma=0.86$). Their return fractions agree to $0.1\%$, so
diffusive similarity holds. Their $\aeff$ curves (Fig.~\ref{fig:alphaeff})
coincide at large scale and fan out below $\sim2\lstar$, ordered by $\gamma$; the
HG--two-term gap reaches $0.15$--$0.22$ near $\lstar$ against a walker-bootstrap
width $\sigma\lesssim0.008$ there, a separation of order twenty standard errors.
The intrinsic far-field endpoint is fixed at $1$ by Eq.~\eqref{eq:alpha} rather
than fitted---the shared large-scale overshoot is the common cutoff artifact
above---so $\aeff$ is a one-parameter probe of the phase function whose anchor is
a theorem, not a free endpoint.

\subsection{How far past $\gamma$ does the index see?}
\label{sec:beyondgamma}

Whether $\aeff$ carries information past the second similarity parameter is the
question Naglic et al.~\cite{naglic2017} posed for subdiffuse reflectance, and
which motivated $\sigma$~\cite{bodenschatz2016} and
$p_{sb}$~\cite{post2020} as third parameters. We test it by matching both $g_1$
and $g_2$, so $\gamma=1.700$ and $\lstar=5$ are fixed, and varying only the
backscatter lobe. Three phase functions serve: a Gegenbauer-kernel form
(Reynolds--McCormick~\cite{reynolds1980}) and two two-term HG mixtures, one with
both lobes forward (MIX-F) and one with a small backward lobe (MIX-B). They
differ in $P(\cos\theta<-0.5)$ by about $10\%$.

The answer is a weak yes, and it takes replication to state it. Across eight
independent seeds per phase function at $5\times10^5$ walkers each, the three
curves agree to better than $0.012$ at every scale above $\lstar$, so $g_2$ does
the work over the experimentally accessible range. At $0.4\,\lstar$ the
seed-averaged indices are $0.489\pm0.004$ (Gegenbauer), $0.509\pm0.002$ (MIX-F)
and $0.487\pm0.007$ (MIX-B): MIX-F separates from the other two by about $0.021$,
roughly three standard errors, while Gegenbauer and MIX-B are indistinguishable.
The direction is the one expected---least backscatter, highest index---but the
three-way ordering is not resolved. A single seed is not enough to see this: the
seed-to-seed standard deviation at $0.4\,\lstar$ is $0.011$ on the extended grid
and $0.026$ on the unextended one, and single-seed spreads across our eight seeds
run from $0.017$ to $0.104$. We therefore report structure past $\gamma$ as
present but small, confined to separations of order two mean free paths where
single and double scattering dominate, and an order of magnitude below the
matched-$g_1$ separation of $0.15$--$0.22$ that $g_2$ itself produces near
$\lstar$.

One further caution. The mixture and Gegenbauer families constrain $g_3$ tightly
once $g_1,g_2$ are fixed---scanning the two-term family at $g_1=0.8$, $g_2=0.66$
gives $g_3\in[0.541,0.560]$, with the Gegenbauer form at $0.5521$ inside
it---so this test varies the backscatter lobe shape more than it varies $g_3$
itself.

A kernel parameterized on $g$ alone is exact for HG and undefined otherwise. In
the subdiffuse regime $g_1$ does not determine the reflectance, and the
literature has reached for $\gamma$~\cite{bevilacqua1999},
$\sigma$~\cite{bodenschatz2016}, and $p_{sb}$~\cite{post2020} in turn. Any
$g$-only kernel carries that scope condition.

\section{Discussion}
\label{sec:discussion}

The result of this paper is a derivation route for the far field. First return
supplies a one-sided $\tfrac12$-stable time; the return-conditioned Gaussian
displacement, subordinated to it, gives the isotropic $1$-stable scaling limit
and the $r^{-3}$ PSF tail, Eq.~\eqref{eq:alpha}; absorption truncates the
subordinator and reproduces the screened radial form of a Farrell source term,
Eq.~\eqref{eq:farrellform}. The index is a theorem, not a fit, which is what lets
$\aeff$ serve as a phase-function probe anchored at $\alpha=1$.

Two things this route does not yet deliver, in increasing difficulty. It
reproduces one screened source, not the two-source finite-depth dipole with its
boundary coefficients. And it says
nothing inside $\lstar$, where the conditional Gaussianity fails
(Fig.~\ref{fig:gauss}) and the near-source excess of Table~\ref{tab:farrell}
lives---both the dipole and the far-field stable law fall short of the Monte
Carlo there. A boundary-kernel correction $K(g,\ns)$, tested by binning the
reflectance restricted to small $\ns$, is the natural next step, and the boundary
truncation factor of \cite{zeller2026cauchy} is a candidate for it: that kernel
is itself Cauchy in $\ns$, empirically, for $g\lesssim2/3$. Whether the two
Cauchy forms are the same object seen from the axial and the lateral side is
open, and it is the question we would ask first.

This places the work against Liemert and Kienle~\cite{liemert2013}, who solve
the anisotropic transport equation for the semi-infinite medium exactly, though
numerically. We do not claim a competing closed form inside $\lstar$. We claim a
first-passage account of why the far-field PSF is Cauchy with index exactly $1$,
and of the screened structure it shares with the diffusion dipole.

\section{Conclusion}
\label{sec:conclusion}

The far-field PSF of a normally incident HG walk is an isotropic $1$-stable law.
Its index is fixed at unity by the subordination of the return-conditioned
Gaussian displacement to a $\tfrac12$-stable first-return time---the product of
the central limit theorem and the distribution-free first-return exponent, which
we verify numerically for the correlated walk. The exit density is Cauchy at
large radius and has no finite second moment, so quantiles, not $\langle
r^2\rangle$, characterize it. Keeping the launch-depth factor that the tail
calculation discards fixes the core radius of that law at
$r_c=\sqrt{2\pi a}\,A_f\simeq1.7\,\lstar$, with the survival amplitude and the
conditional variance as its only inputs; two independent readings of the
simulations agree with it to about $10\%$.
Absorption weights each path by $e^{-\mu_a L}$ and
yields the screened radial form that each source term of the Farrell dipole
carries, though not the full finite-depth two-source construction. A
scale-resolved index reads the phase function off the reflectance and tracks
$\gamma$ down to about one transport mean free path.

One question remains open and testable: whether a boundary kernel $K(g,\ns)$
captures the near-source excess of Table~\ref{tab:farrell}, which can be checked
by binning $R(\rho)$ on small $\ns$ against the kernel directly. The
matched-$(g_1,g_2)$ test of Sec.~\ref{sec:beyondgamma} is settled to the accuracy
we can reach: replicated over eight seeds, the separation past $\gamma$ survives
at about three standard errors below $0.5\,\lstar$, an order of magnitude smaller
than the separation $g_2$ itself produces near $\lstar$, and it does not resolve
an ordering among the three phase functions.

\section*{Acknowledgments}

We thank the anonymous referee, whose reports sharpened the scope of the claims
in Secs.~\ref{sec:core}, \ref{sec:tempered} and \ref{sec:beyondgamma}.

Drafting and revision of the manuscript text used a large language model
(Claude, Anthropic). The physics, the Monte Carlo kernel, the runs and the
analysis are our own; no result, figure, table or reference was produced by the
tool. We are responsible for the content of the paper, including any error the
tool failed to catch.

\appendix
\section{Reproducing the runs}
\label{sec:repro}

Every result here comes from one Monte Carlo kernel and four reductions of its
output. The kernel advances all live walkers together: it draws exponential free
paths, rotates the direction by the HG angle, and records $\ns$, the exit point
$(x,y)$ by linear interpolation along the crossing flight, the total path length
$L=\sum_i s_i$ for the absorption weight $e^{-\mu_a L}$, and the censored count.
It updates about $6.5\times10^6$ walker-steps per second on one core; $10^6$
walkers at $g=0.8$ took 113~s. The reductions are the absorption reweighting and
Farrell comparison of Sec.~\ref{sec:tempered}, the matched-$g_1$ index sweep of
Sec.~\ref{sec:alphaeff}, the matched-$(g_1,g_2)$ sweep of
Sec.~\ref{sec:beyondgamma} repeated over independent seeds, and the
flight-ceiling sweep of Fig.~\ref{fig:ceiling}. For $10^8$ walkers, about 3
core-hours at $g=0.8$, run one process per core with independent seeds and merge
histograms. A deterministic alternative for a noise-free $K(g,\ns)$ is a
$(z,\mu)$ iteration at fixed transverse wavevector, parallel over $k$.

The binned first-return and exit-radius counts behind Figs.~\ref{fig:firstreturn},
\ref{fig:survival} and~\ref{fig:psf}, at $g=0$, $0.5$, $0.8$ and $0.9$ with the
per-$g$ walker and censored counts, are available from the authors; they
reproduce the return fractions of Sec.~\ref{sec:survival} and the slopes quoted
above. Figures~\ref{fig:gauss}, \ref{fig:alphaeff} and~\ref{fig:ceiling} need the
per-walker $(x,y,\ns,L)$ samples, which we have not archived.

\paragraph{A validation blind spot at $g=0$.}
An early implementation had a threshold in the direction-rotation kernel that
made the $|u_z|\to1$ frame degenerate: a walker landing on the pole stayed there
and the walk collapsed to one dimension. The $g=0$ run still returned a
first-return slope of $-1.517$, correct to the eye. It passed because
Sparre~Andersen universality (Sec.~\ref{sec:firstreturn}) holds for any
symmetric continuous one-dimensional walk, so a broken lateral kernel is
invisible to it. Only $g=0.8$, slope $-1.333$ through the naive window, exposed
the fault. A validation gate built on $g=0$ inherits this blind spot; test the
lateral kernel at $g>0$, or test isotropy of the scatter operator directly.

Sanity checks worth rerunning after any edit: sampled $\langle\cos\theta\rangle$
against $g$; isotropy of the scatter operator applied at the pole
($\langle v_z\rangle=0$, $\langle v_z^2\rangle=1/3$ at $g=0$); unit norm of the rotated direction; the $g=0.8$ first-return slope, which is what
caught the pole bug of Sec.~\ref{sec:model}; and the $g=0$ survival amplitude
under an isotropic first flight conditioned to $\zeta_1>0$, which
Sparre~Andersen fixes at $2/\sqrt\pi$ exactly. The last of these is the only
check in the list that tests an absolute number rather than an exponent or a
ratio.

\paragraph{Definitions and uncertainties.} Radial densities use the areal
convention of Fig.~\ref{fig:psf}; first-return densities divide counts by the
logarithmic bin width and the walker count. Censored walkers are
excluded from every density and carry no exit point. Absorption normalization is
by the same non-absorbing walker count $N$, so the integrated PSF is the total
diffuse reflectance and equals the summed weight $\sum e^{-\mu_a L_i}/N$. The
non-HG phase functions are two-term HG, $w\,\mathrm{HG}(g_f)+(1-w)\,\mathrm{HG}
(-g_b)$; $\delta$-plus-isotropic, $w\,\delta(\cos\theta-1)+(1-w)/2$; and the
Gegenbauer (Reynolds--McCormick) kernel, sampled by the inverse CDF of
$(1+g_k^2-2g_k\cos\theta)^{-(a+1)}$. Table~\ref{tab:phasefns} lists every weight
and component parameter with the resulting moments. Quoted percentages are
seed-to-seed empirical variation where a second
seed was run, and Poisson counting error
$\sqrt{H}/H$ per bin otherwise; slope fits are least squares over the stated
window. The seed-to-seed standard deviation of $\aeff$ is $\lesssim0.005$ on the
plateau and $0.011$ at $0.4\,\lstar$ on the extended grid, rising to $0.026$
there if the grid is not extended and the endpoint difference is one-sided. The
matched-$(g_1,g_2)$ comparison of Sec.~\ref{sec:beyondgamma} is averaged over
eight independent seeds per phase function; quoted uncertainties on those indices
are standard errors of that mean. Runs differ in flight ceiling: $\ns\le10^4$ for
Figs.~\ref{fig:firstreturn}--\ref{fig:psf} and Table~\ref{tab:farrell},
$\ns\le10^5$ for the plateau indices of Sec.~\ref{sec:alphaeff}, and
$10^4$, $10^5$, $10^6$ for the three curves of Fig.~\ref{fig:ceiling}.

\begin{table}[htbp]
  \centering
  \caption{Every plotted phase function, with its parameters and the first two
    Legendre moments $g_1,g_2$ and similarity parameter
    $\gamma=(1-g_2)/(1-g_1)$. The upper block is the matched-$g_1$ set of
    Sec.~\ref{sec:alphaeff}; the lower is the matched-$(g_1,g_2)$ set of
    Sec.~\ref{sec:beyondgamma}. All share $g_1=0.800$ and $\lstar=5$.}
  \label{tab:phasefns}
  \begin{tabular}{llccc}
    \toprule
    label & definition & $g_1$ & $g_2$ & $\gamma$ \\
    \midrule
    HG           & $\mathrm{HG}(0.800)$ & 0.800 & 0.640 & 1.80 \\
    $\delta$+iso & $0.80\,\delta(\cos\theta-1)+0.20\cdot\tfrac12$ & 0.800 & 0.800 & 1.00 \\
    two-term HG  & $0.90\,\mathrm{HG}(0.9444)+0.10\,\mathrm{HG}(-0.500)$ & 0.800 & 0.828 & 0.86 \\
    \midrule
    GK           & $(1+g_k^2-2g_k\cos\theta)^{-(a+1)},\ a=0.30141,\ g_k=0.87218$ & 0.800 & 0.660 & 1.70 \\
    MIX-F        & $0.50\,\mathrm{HG}(0.94142)+0.50\,\mathrm{HG}(0.65858)$ & 0.800 & 0.660 & 1.70 \\
    MIX-B        & $0.98\,\mathrm{HG}(0.82020)+0.02\,\mathrm{HG}(-0.18995)$ & 0.800 & 0.660 & 1.70 \\
    \bottomrule
  \end{tabular}
\end{table}

\end{document}